\documentclass[a4paper,preprint,superscriptaddress,preprintnumbers,nofootinbib]{revtex4}

\usepackage{amsmath}
\usepackage{comment}
\usepackage[dvipdfmx]{graphicx}

\usepackage{dcolumn}
\usepackage{bm}
\usepackage[dvipdfmx]{color}
\usepackage[normalem]{ulem}

\allowdisplaybreaks

\newcommand{\be}{\begin{eqnarray}}
\newcommand{\ee}{\end{eqnarray}}
\catcode`\@=11
\def\lsim{\mathrel{\mathpalette\@versim<}}
\def\gsim{\mathrel{\mathpalette\@versim>}}
\def\@versim#1#2{\vcenter{\offinterlineskip
\ialign{$\m@th#1\hfil##\hfil$\crcr#2\crcr\sim\crcr } }}
\catcode`\@=12
\newcommand{\del}{\partial}

\newcommand{\al}[1]{\begin{align}#1\end{align}}
\newcommand{\bp}{\begin{pmatrix}}
\newcommand{\ep}{\end{pmatrix}}
\newcommand{\nn}{\nonumber\\}

\newcommand{\p}{\partial}

\newcommand{\bs}[1]{\boldsymbol}

\newcommand{\pmat}[1]{\begin{pmatrix}#1\end{pmatrix}}

\newcommand{\fn}[1]{\!\left(#1\right)}

\newcommand{\bra}{\langle}
\newcommand{\ket}{\rangle}
\newcommand{\Lag}{\mathcal L}

\begin{document}
\title{Scale genesis and gravitational wave 
in a classically scale invariant extension of the standard model}

\author{Jisuke \surname{Kubo}}
\affiliation{Institute for Theoretical Physics, Kanazawa University, Kanazawa 920-1192, Japan}

\author{Masatoshi \surname{Yamada}}
\affiliation{Department of Physics, Kyoto University, Kyoto 606-8502, Japan\\
Institut f\"ur Theoretische Physik, Universit\"at Heidelberg, Philosophenweg 16, 69120 Heidelberg, Germany
}

\preprint{
KANAZAWA-16-11
}
\preprint{
KUNS-2643
}

\begin{abstract}
We assume that the origin of the electroweak (EW) scale
is a gauge-invariant scalar-bilinear condensation
in a strongly interacting non-abelian gauge  sector,
which is connected to the standard model
via a Higgs portal coupling.
The dynamical scale genesis appears as a phase transition
at finite temperature, and it can  produce
a gravitational wave (GW)  background in the early Universe.
We find that 
the critical temperature of the scale phase transition
lies above that of the EW
phase transition
and below few $O(100)$ GeV and it is strongly first-order.
We calculate the spectrum of  the  GW background
and find
the scale phase transition is strong enough that 
the   GW background   
can be observed by DECIGO.

\end{abstract}
\maketitle

\newpage
\section{Introduction}
One of the problems of the standard model (SM) is that it
cannot explain the origin of the electroweak (EW) scale.
Obviously, if we start with a theory which contains a scale from the 
beginning, we have no chance to explain its origin.
We recall that 
the Higgs mass term is the only term that breaks scale invariance at the classical level in the SM. 
So, a central question is: What is the origin of the Higgs mass term?
Recently, a number of studies on 
a scale invariant extension of the SM have been performed, where two ideas
have been considered.
The one \cite{Fatelo:1994qf,Hempfling:1996ht,Hambye:1995fr,Meissner:2006zh,Foot:2007as,Foot:2007ay,Chang:2007ki,Hambye:2007vf,Iso:2009ss,Holthausen:2009uc,Ishiwata:2011aa,Englert:2013gz,Khoze:2013oga,Carone:2013wla,Farzinnia:2013pga,Gretsch:2013ooa,Kawamura:2013kua,Hambye:2013sna,Khoze:2013uia,Gabrielli:2013hma,Abel:2013mya,Ibe:2013rpa,Hill:2014mqa,Guo:2014bha,Radovcic:2014rea,Khoze:2014xha,Salvio:2014soa,
Davoudiasl:2014pya,Chankowski:2014fva,Allison:2014zya,Farzinnia:2014xia,Ko:2014loa,Altmannshofer:2014vra,Kang:2014cia,Giudice:2014tma,Guo:2015lxa,Kannike:2015apa,Ahriche:2015loa,Haba:2015lka,  Plascencia:2015xwa,Khoze:2016zfi,Karam:2015jta,Karam:2016rsz} relies on the Coleman-Weinberg (CW) potential
\cite{Coleman:1973jx}, and
the other one is based on non-perturbative effects:
 Dynamical chiral symmetry breaking
\cite{Nambu:1960xd,Nambu:1961tp} is applied in \cite{Hur:2007uz,Hur:2011sv,Heikinheimo:2013fta,Holthausen:2013ota,Kubo:2014ova,Kubo:2014ida,Antipin:2014qva,Heikinheimo:2014xza,Carone:2015jra,Ametani:2015jla,Haba:2015qbz,Kannike:2016bny,Hatanaka:2016rek}  and gauge-invariant  scalar-bilinear  condensation 
\cite{Fradkin:1978dv,Abbott:1981re,Chetyrkin:1982au} is used in \cite{Kubo:2015cna,Kubo:2015joa}.

In the case that a scale is dynamically created, there will be a corresponding scale phase transition.
Chiral phase transition is a scale phase transition in this context.
If a scale phase transition is strongly first-order, the Universe undergoes
a strong phase transition  at a certain 
 high temperature, thereby producing  gravitational wave (GW) background 
that could be observed today 
\cite{Witten:1984rs,Hogan:1986qda,Turner:1990rc}. There are a number of 
GW experiments that are on-going or planned in the near and far future
(see  \cite{Maggiore:1999vm,Binetruy:2012ze} and also 
\cite{Kahniashvili:2008pf,Caprini:2015zlo,Schwaller:2015tja}
for instance). It is in fact this year in which LIGO 
 \cite{Abbott:2016blz} has observed the GW for the first time.

In this paper we focus on the second non-perturbative effect,
the gauge-invariant  scalar-bilinear  condensation, that produces
the GW background in the early Universe.
The present work is a natural extension of our recent works
\cite{Kubo:2015cna,Kubo:2015joa}, because we have shown there that the scale phase transition
due to the scalar-bilinear condensation
is strongly first-order in a wide region  of the parameter space.
We have performed the analysis, using an effective field theory
that describes approximately the process of the dynamical 
scale genesis via the scalar-bilinear condensation.
Since we will be using the same effective field theory
to calculate the spectrum of  the corresponding GW background,
we devote, for completeness, first two chapters for briefly elucidating
our effective field theory method, where 
parallels to the 
Nambu-Jona-Lasinio (NJL) theory \cite{Nambu:1961tp} could be read off.
In \cite{Kubo:2015cna,Kubo:2015joa} 
we have used the so-called self-consisting mean-field
approximation~\cite{Kunihiro:1983ej,Hatsuda:1994pi} 
to derive the mean-field Lagrangian, 
we will employ the path-integral approach \cite{Hubbard:1959ub,Stratonovich:1957} (sect.~\ref{models}) to arrive
at the same Lagrangian, a procedure which  may be more 
readable in high-energy physics society.

We  will single out a benchmark point in the parameter space
in section~\ref{explicit point}. 
The  benchmark point represents
the wide region of the parameter space, which
is consistent with the dark matter (DM) phenomenology.
(In the model we will consider there exists a DM candidate
due to an unbroken flavor symmetry in the hidden sector.)
In this region of the parameter space the scale phase transition is
strongly first-order.
 
In section~\ref{GW analysis} we will calculate the spectrum of  the corresponding GW background.
There are three production mechanisms of  GWs at 
a strong first-oder phase transition,
in which the bubble nucleation grows and the GW is produced;
collisions of bubble walls  $\Omega_\text{coll} $ \cite{Kosowsky:1991ua,Kosowsky:1992rz,Kosowsky:1992vn,Kamionkowski:1993fg,Caprini:2007xq,Huber:2008hg,Jinno:2016vai},
magnetohydrodynamic (MHD) turbulence $\Omega_\text{MHD}$
\cite{Kosowsky:2001xp,Dolgov:2002ra,Caprini:2006jb,Gogoberidze:2007an,Kahniashvili:2008pe,Caprini:2009yp,Kisslinger:2015hua}
and also
sound waves $\Omega_\text{sw}$
after the bubble wall collisions \cite{Hindmarsh:2013xza,Hindmarsh:2015qta,Giblin:2013kea,Giblin:2014qia}.
Using the formulas given in these papers and especially in 
\cite{Caprini:2015zlo}, we will compute  these individual 
contributions to the GW background spectrum 
for a set of the benchmark parameters
and find that
$\Omega_\text{coll}$ and
$\Omega_\text{MHD}$ are several orders of magnitude
smaller than
$\Omega_\text{sw}$.  
Finally we will compare our result with the sensitivity
of various GW experiments. We will find that 
the scale phase transition caused by the scalar-bilinear
condensation can be strong enough to  produce
the  GW background that
can be observed by DECIGO \cite{Seto:2001qf}.

Sect.~\ref{summary} is devoted to a summary, and in the appendix
we compute the field renormalization factor.

\section{The basic idea and the path-integral approach}\label{models}
\noindent 
We consider a classical scale invariant extension of the 
SM, which has been studied in \cite{Kubo:2015cna,Kubo:2015joa}.
The basic assumption there is that the origin of 
the  EW scale  is  a scalar-bilinear condensation, which forms due a strong
non-abelian gauge interaction in a hidden sector and 
triggers the EW symmetry breaking through a Higgs portal coupling.
The hidden sector is described by
an $SU(N_c)$ gauge theory with
 the scalar fields $S_i^{a}~
(a=1,\dots,N_c,i=1,\dots,N_f)$ in
the fundamental representation of $SU(N_c)$.
Accordingly, the total Lagrangian is given by
\al{
{\cal L}_{\rm H} &=
-\frac{1}{2}~\mbox{tr} F^2
+([D^\mu S_i]^\dag D_\mu S_i)
-\hat{\lambda}_{S}(S_i^\dag S_i) (S_j^\dag S_j)\nn
&\quad -\hat{\lambda'}_{S}
(S_i^\dag S_j)(S_j^\dag S_i)
+\hat{\lambda}_{HS}(S_i^\dag S_i)H^\dag H
-\lambda_H ( H^\dag H)^2+{\cal L}'_{\mathrm{SM}},
\label{LH}
}
where $D_\mu S_i = \partial_\mu S_i -ig_H G_\mu S_i$, $G_\mu$ 
is the matrix-valued $SU(N_c)$ gauge field,  and
the SM Higgs doublet field is denoted by $H$
(the parenthesis stands for $SU(N_c)$ invariant products.).
The last term, ${\cal L}'_{\mathrm{SM}}$, 
contains the SM gauge and Yukawa interactions.
Note that the Higgs mass term, which is the only
scale-invariance violating term in the SM, is absent in \eqref{LH}.

We  assume that the $SU(N_c)$ gauge theory in the hidden sector
 is asymptotically free
and gauge symmetry is unbroken for the entire
energy scale, such that above  a certain 
energy the theory is perturbative and there is no mass scale 
except for  renormalization scale, which is present
due to  scale anomaly \cite{Callan:1970yg}.
At a certain low energy scale (say the confinement scale)
the gauge coupling $g_H$ becomes so large that
the $SU(N_c)$ invariant scalar bilinear dynamically
forms  a $U(N_f)$ invariant
condensate \cite{Abbott:1981re,Chetyrkin:1982au},
\al{
\langle (S^\dag_i S_j)\rangle =
\langle \sum_{a=1}^{N_c} S^{a\dag}_i S^a_j~\rangle\propto \delta_{ij},
\label{condensate}
} 
and the mass term (constituent mass) for $S_i$ is dynamically  generated.
The creation of the mass term from nothing
is possible only by a  non-perturbative effect:
Scale anomaly  \cite{Callan:1970yg} cannot generate a mass term,
because the $SU(N_c)$ gauge symmetry is  unbroken
by assumption.
Of course, scale anomaly does contribute
to the mass once it is generated, and  how it logarithmically runs
is  described, for instance in \cite{Aoki:2012xs}, in a modern language of renormalization\footnote{There exist proofs within the framework of perturbation theory
that conformal anomaly 
does not generate  mass term of scalar field
in the class of  massless renormalizable (not super-renormalizable)
field theories. This  is rigorously proven  in the massless $\phi^4$ theory 
by Loewenstein and Zimmermann \cite{Lowenstein:1975rf}, and 
the same conclusion
was made  in massless non-abelian gauge theories  in \cite{Poggio:1976qr}.
If the regularization (e.g. cut-off regularization) 
 breaks scale invariance, one needs 
a counter term for the mass of the scalar field.
From this reason we will employ  dimensional regularization.
}.

It has been intended in \cite{Kubo:2015cna,Kubo:2015joa} to describe the non-perturbative
phenomena of condensation \eqref{condensate} approximately by using an effective theory.
That is, the effective theory should  describe the dynamical generation of the
mass for $S_i$ via the scalar-bilinear condensation \eqref{condensate}.
Using the effective theory it should be also possible to approximately 
describe how the energy scale transfers from the hidden sector to the SM sector.
Inspired by the NJL theory, which can approximately describe the dynamical
chiral symmetry breaking in QCD, we have assumed that the effective Lagrangian 
does not contain the $SU(N_c)$ gauge fields, because 
they are integrated out, while it contains
the ``constituent'' scalar fields $S_i^{a}$.
Furthermore, since the hard breaking of scale invariance by anomaly  is only logarithmic, 
the non-perturbative  breaking may be assumed to be dominant,
 so that  scale anomaly may be ignored
 in writing down  an effective Lagrangian at the tree level:
 \al{
  {\cal L}_{\rm eff} &=
 ([\partial^\mu S_i]^\dag \partial_\mu S_i)-
\lambda_{S}(S_i^\dag S_i) (S_j^\dag S_j)
-\lambda'_{S}
(S_i^\dag S_j)(S_j^\dag S_i)\nn
&\quad +\lambda_{HS}(S_i^\dag S_i)H^\dag H
-\lambda_H ( H^\dag H)^2,
\label{Leff}
}
where   all $\lambda$'s are positive. 
This is the most general form which is
consistent with 
the global $SU(N_c)\times U(N_f)$ symmetry and the classical scale invariance
(we have suppressed the kinetic term for $H$ in \eqref{Leff}
because it does not play any important role for our 
discussions here).
That is, ${\cal L}_{\rm eff}$ 
has the same global symmetry as  ${\cal L}_{\rm H}$ 
even at the quantum level.
We emphasize  that the effective Lagrangian \eqref{Leff}
is not a Lagrangian for an effective theory of \eqref{LH} after the confinement 
has taken place and the condensation \eqref{condensate} has appeared:
It should describe the process of the condensation.
Note also that the couplings
$\hat{\lambda}_{S}, \hat{\lambda'}_{S}$, and $\hat{\lambda}_{HS}$ in ${\cal L}_{\rm H}$ are not the same as
$\lambda_{S}, \lambda'_{S}$, and $\lambda_{HS}$ in $ {\cal L}_{\rm eff}$,
because the latter are effective couplings which are dressed by
 hidden gluon contributions.
 
\subsection{Path-integral formalism}\label{path integral}
In \cite{Kubo:2015cna,Kubo:2015joa} the self-consistent mean-field approximation \cite{Kunihiro:1983ej,Hatsuda:1994pi} has been applied to treat the effective Lagrangian \eqref{Leff}.
Here we base on  the path-integral formalism
to obtain the mean-field Lagrangian ${\cal L}_{\rm MFA}$.
The method is known as the so-called auxiliary field method or Hubbard--Stratonovich transformation \cite{Hubbard:1959ub,Stratonovich:1957}.
The starting path-integral with the effective Lagrangian \eqref{Leff} is given by
\al{
Z=\int \mathcal{D} S^\dag \mathcal{D} S \, \exp\bigg[ i\int d^4x\,& \bigg\{ ([\partial^\mu S_i]^\dag \partial_\mu S_i)-
\lambda_{S}(S_i^\dag S_i) (S_j^\dag S_j)
-\lambda'_{S}
(S_i^\dag S_j)(S_j^\dag S_i)\nn
&\quad +\lambda_{HS}(S_i^\dag S_i)H^\dag H
-\lambda_H ( H^\dag H)^2 \bigg\}  \bigg],
\label{pathintegral1}
}
where we focus  only on the path-integral for $S^\dag$ and $S$.
We insert the Gaussian integral 
\al{
1 &\propto  \int {\cal D} f'\,{\cal D} \phi'
\exp \left[ i \int d^4 x\,{\cal L}_A  \right]&
&\text{with}&
{\cal L}_A &=N_f (N_f\lambda_S+\lambda'_S)f'^2
+\frac{\lambda'_S}{2}\phi'^a \phi'^a
}
to the path-integral~\eqref{pathintegral1} and
make the change of the integration variables to $f$ and $\phi^a_0$
according to
\al{
f'&=f-(S_i^\dag S_i)/N_f,& \phi'{}^a&=\phi^a_0-2 (S_i^\dag t^a_{ij} S_j),&
}
where $t^a~(a=1,\dots, N_f^2-1)$ are the $SU(N_f)$ generators
in the fundamental representation.
Then the path-integral can be written as 
\al{
Z&=
\int \mathcal{D} S^\dag \mathcal {D} S{\cal D} f{\cal D} \phi _0\, \exp\bigg[ i\int d^4x\,\bigg\{
 ([\partial^\mu S_i]^\dag \partial_\mu S_i) 
 +\lambda_{HS}(S_i^\dag S_i)H^\dag H
 -\lambda_H ( H^\dag H)^2 \nn
&\quad + N_f (N_f\lambda_S+\lambda'_S) f^2 
-2(N_f\lambda_S+\lambda'_S)f (S_i^\dagger S_i)
+\frac{\lambda_S'}{2}(\phi^a_0)^2
-2\lambda_S' \phi^a_0(S_i^\dagger t^a_{ij} S_j)
\bigg\}
\bigg],
\label{pathintegral2}
}
where we have used the identity
\al{
(S_i^\dagger t^a_{ij} S_j)(S_k^\dagger t^a_{kl} S_l)
&=S_i^\dagger S_jS_k^\dagger S_l \left(  t^a_{ij}t^a_{kl}  \right)\nn
&= S_i^\dagger S_jS_k^\dagger S_l\times \frac{1}{2}\left( \delta_{il}\delta_{jk}-\frac{1}{N_f}\delta_{ij}\delta_{kl} \right)\nn
&=\frac{1}{2}\left( 
(S_i^\dagger S_j)(S_j^\dagger S_i) -\frac{1}{N_f} (S_i^\dagger S_i)^2
\right).
}
Note that the Euler--Lagrange equations for 
the auxiliary  fields $f$ and $\phi^a_0$ become $f=(S_i^\dagger S_i)/N_f$ and $\phi^a_0=2(S_i^\dagger t^a_{ij} S_j)$, respectively, and substituting them into \eqref{pathintegral2}, we are back to \eqref{pathintegral1}.
We thus arrive at  the mean-field Lagrangian
\al{
\Lag_\text{MFA}&=
 ([\partial^\mu S_i]^\dag \partial_\mu S_i) 
 -M^2(S_i^\dag S_i)
 -\lambda_H ( H^\dag H)^2 \nn
&\quad + N_f (N_f\lambda_S+\lambda'_S) f^2 
+\frac{\lambda_S'}{2}(\phi^a_0)^2
-2\lambda_S' \phi^a_0(S_i^\dagger t^a_{ij} S_j),
\label{MFA2ap}
}
where 
\al{
M^2 = 2(N_f\lambda_S+\lambda'_S)f -\lambda_{HS} H^\dagger H.
\label{cons scala mass}
}
If we expand the composite field $f$ and the Higgs doublet around the 
vacuum values $f_0$ and $v_h$, i.e.,
\al{
f&=f_0 + Z_\sigma^{1/2} \sigma,&
H&= \frac{1}{\sqrt{2}}
\pmat{
\chi^1+i\chi^2\\
v_h+h +i\chi^0
},
\label{field expansion}
}
and redefine $\phi^a_0$ as $\phi^a_0=Z_\phi^{1/2}\phi^a$,
the mean-field Lagrangian \eqref{MFA2ap} becomes
\al{
\Lag_\text{MFA}'&=
 ([\partial^\mu S_i]^\dag \partial_\mu S_i) 
 -M_0^2(S_i^\dag S_i)
 + N_f (N_f\lambda_S+\lambda'_S)Z_\sigma \sigma^2
 +\frac{\lambda'_S}{2} Z_\phi \phi^a\phi^a \nn
 &\quad-2 (N_f\lambda_S+\lambda'_S)Z_\sigma^{1/2} \sigma (S_i^\dagger S_i)
-2\lambda'_S Z_\phi^{1/2}(S_i^\dagger t^a_{ij}\phi^a S_j)\nn
&\quad +\frac{\lambda_{HS}}{2}(S_i^\dag S_i) h (2v_h +h)
-\frac{\lambda_H}{4}h^2(6v_h^2 +4v_h h + h^2),
\label{MFA3}
 }
 where $Z_\sigma$ and $Z_\phi$ are field renormalization constants
 of the canonical dimension two.
In \eqref{MFA3}  we have neglected the would-be Goldstone bosons $\chi^i$ in the Higgs  field and defined the constituent scalar mass squared as
\al{
M_0^2 = 2(N_f \lambda_S +\lambda'_S)f_0 -\frac{\lambda_{HS}}{2}v_h^2.
}

\subsection{Effective potential and the mean-field vacuum}\label{effective potential subsection}
To determine the mean-field vacuum, we next derive the mean-field effective potential by integrating out the quantum fluctuations of $S$.
The assumption that the bilinear condensate \eqref{condensate}  is $U(N_f)$ invariant means  for  the composite fields that
$ \bra f\ket \neq 0$ and $\bra \phi^a \ket =0$.
Therefore, it is sufficient to  consider the path-integral \eqref{MFA2ap} with $\phi^a=0$.
Since \eqref{MFA2ap} is quadratic in $S$, the path-integral for $S$ is 
 Gaussian, and then we obtain\footnote{
The field $S$ is expanded around the homogeneous background, i.e., $S\to \bar S +\chi$, and the path-integral of $\chi$ is performed.
}
\al{
Z&=\int {\cal D} f\, \exp\bigg[ i\int d^4x\, \bigg\{
-M^2(\bar S^\dagger \bar S)
 -\lambda_H ( H^\dag H)^2\nn
&\quad+ N_f (N_f\lambda_S+\lambda'_S) f^2 
 + iN_cN_f\ln \text{Det}\left[ \p^2 + M^2 \right]
 \bigg\}
\bigg],
\label{pathintegral3}
}
where the fluctuation of $S$ has been integrated out 
around its background field $\bar S$.
The last term in the right-hand side of \eqref{pathintegral3} 
 is evaluated as
\al{
\ln \text{Det}\left[ \p^2 + M^2 \right]
 = (VT) \frac{M^4}{2(4\pi)^2}\left( \frac{1}{\bar \epsilon} -\ln\fn{M^2} +\frac{3}{2} \right),
}
where the dimensional regularization has been used, 
$VT$ is the space-time volume ($VT=\int d^4x$), and $1/{\bar \epsilon} = 2/(d-4) -\gamma_E +\ln\fn{4\pi}$.
Then the 1PI effective action at the one-loop level is given by
\al{
\Gamma[\bar S,f,H]&= VT\bigg[
- M^2(\bar S^\dagger \bar S)
-\lambda_H ( H^\dag H)^2 \nn
&\quad + N_f (N_f\lambda_S+\lambda'_S) f^2 
+  \frac{N_cN_f}{2(4\pi)^2}M^4\left( \frac{1}{\bar \epsilon} -\ln\fn{M^2} +\frac{3}{2} \right)
 \bigg].
 \label{EAgeneral}
}
Note that we have not used the large-$N$ approximation to derive the effective action \eqref{EAgeneral}.
Finally we  obtain the effective potential
\al{
V_\text{MFA}\fn{\bar S,f,H}
&=-\frac{\Gamma[\bar S,f,H]}{VT}\nn
&= M^2 (\bar S_i^\dagger \bar S_i) +\lambda_H (H^\dagger H)^2 -N_f (N_f \lambda_S +\lambda'_S)f^2 +\frac{N_cN_f}{32\pi^2}M^4 \ln\frac{M^2}{\Lambda_H^2},
\label{effective potential MFAa}
}
where the divergence $1/{\bar \epsilon}$ was removed by  renormalization of the coupling constants in the  $\overline{MS}$ scheme and $\Lambda_H = \mu
e^{3/4}$ is the scale at which the quantum correction vanishes
if $M=\Lambda_H$. 
Note here that the scale $\Lambda_H$ is generated by quantum effect in the 
classically scale invariant effective theory \eqref{Leff} and 
becomes the origin of the electroweak scale as it will be seen below.

The  minima of the effective potential \eqref{effective potential MFAa} 
can be obtained from  the solution of the gap equations\footnote{A similar potential problem has been studied in 
\cite{Coleman:1974jh,Kobayashi:1975ev,Abbott:1975bn,Bardeen:1983st}.
But they did not study the classical scale invariant case in detail,
and moreover no coupling to the SM was introduced.}
\al{
0=\frac{\del}{\del \bar S^a_i}V_{\rm MFA}
= \frac{\del}{\del f}V_{\rm MFA}
=\frac{\del}{\del H_l}V_{\rm MFA}~(l=1,2).
\label{station}
}
The first equation \eqref{station} yields $\bra \bar S^a_i\ket \bra M^2\ket=0$, which is satisfied in the following three cases: (i)~$\bra \bar S^a_i\ket\neq 0$ and $\bra M^2\ket=0$; (ii)~$\bra \bar S^a_i\ket= 0$ and $\bra M^2\ket=0$; (iii)~$\bra \bar S^a_i\ket= 0$ and $\bra M^2\ket\neq 0$.
The case (i) corresponds to the end-point solution~\cite{Bardeen:1983st} in which the effective potential has a flat direction, i.e., $V_\text{eff}=0$ for $f=H=0$. The gap equations in the case (i) imply the relation
 $\bra f \ket=2\lambda _H/N_f \lambda_{HS}\bra H^\dagger H \ket$, 
 and the effective potential at the minimum 
vanishes, i.e., $\bra V_\text{MFA}\ket = 0$  for an arbitrary value of $\bar S$.
In the case (ii) $\bra V_\text{MFA}\ket =0$ follows trivially.
In the case (iii), using the other gap equations, we obtain 
\al{
 |\langle H\rangle |^2
 &=\frac{v_h^2}{2}=
 \frac{N_f\lambda_{HS}}{G}\Lambda_H^2\exp\left(  \frac{32\pi^2 \lambda_H}{N_c G}-\frac{1}{2}\right),~
 \langle f\rangle =f_0=\frac{2 \lambda_H}{N_f\lambda_{HS}} 
 |\langle H\rangle |^2,
 \label{vev1}\\
 \langle M^2\rangle &= M_0^2=\frac{G}{N_f\lambda_{HS}}
  |\langle H\rangle |^2
  \label{M02}
}
at the minimum, where $G \equiv 4N_f \lambda_H \lambda_S-N_f \lambda_{HS}^2+4 \lambda_H\lambda'_S$.
The value of the effective potential at this minimum is given by
\al{
\bra V_\text{MFA}\ket = -\frac{N_cN_f}{64\pi^2}\Lambda_H^4\exp\left( \frac{64\pi^2\lambda_H}{N_c G}-1\right)<0.
}
We therefore conclude that the case (iii) corresponds 
to  the absolute minimum of the effective potential \eqref{effective potential MFAa}
as far as $G>0$ is satisfied.
The Higgs mass at this level of approximation is calculated to be
\al{
m_{h0}^2= |\langle H\rangle |^2\left(
\frac{16\lambda_H^2
(N_f\lambda_S+\lambda'_S)}{G}
+\frac{N_c N_f
\lambda_{HS}^2}{8\pi^2}\right).
\label{vev2}
}
In the small $\lambda_{HS}$ limit, this mass can be expanded as
\al{
m_{h0}^2=2N_f\lambda_{HS} \bra f\ket 
+N_f\lambda_{HS}^2\left(\frac{N_c}{8\pi^2 } + \frac{1}{N_f \lambda_S +\lambda_S'} \right) |\bra H \ket |^2
+\cdots.
\label{expanded higgs mass}
}
We see that the Higgs mass is generated by the scalar-bilinear condensation
$f$ and the second term in the right-hand side of \eqref{expanded higgs mass} comes from the back-reaction due to the finite vacuum expectation value of the Higgs field.
In other words, the first term is, due to \eqref{vev1},
equal to $4\lambda_H |\langle H\rangle |^2$, which is  the tree level expression in the SM model,
so that the second term must be a correction due to the back-reaction.
The correction coming from the SM sector to the Higgs 
mass \eqref{vev2} will be calculated below.

\subsection{Corrections from the SM sector}
We calculate the one-loop contribution
 from the SM sector 
to the effective potential 
\eqref{effective potential MFAa} and evaluate the correction to the Higgs mass \eqref{vev2}.
The  one-loop contribution to the effective potential can be calculated from
\al{
V_\text{CM}\fn{h} = \sum_{I=W,Z,h} \frac{n_I}{2}\int \frac{d^4k}{(2\pi)^4} \ln\fn{k^2 + m_I^2\fn{h}}
- \frac{n_t}{2}\int \frac{d^4k}{(2\pi)^4} \ln\fn{k^2 + m_t^2\fn{h}}+\mbox{c.t.},
\label{one loop potential}
}
where $n_I$ ($I=W,Z,t,h$) is the degrees of freedom of
the corresponding  particle, i.e., $n_W=6$, $n_Z=3$, $n_t=12$ and $n_h=1$,
and c.t. stands for the counter terms. The contributions from the would-be Goldstone bosons in  the Higgs field have been neglected\footnote{We work in the Landau gauge.}.
As before we use the dimensional regularization to respect scale invariance,
and the counter terms are so  chosen that the  normalization
conditions
\al{
V_{\rm CW}\fn{h=v_h} &=0,&
\frac{d V_{\rm CW}\fn{h} }{d h}\bigg|_{h=v_h}&=0&
\label{normalization}
}
with $v_h=246$ GeV are satisfied. 
In this way  we obtain the Coleman--Weinberg potential~\cite{Coleman:1973jx} 
\al{
V_{\rm CW}\fn{h} &=
C_0 (h^4-v_h^4)+\frac{1}{64 \pi^2}
\left[ ~6 \tilde{m}_W^4 \ln (\tilde{m}_W^2/m_W^2)+
3 \tilde{m}_Z^4 \ln (\tilde{m}_Z^2/m_Z^2)\right.\nn
 &\left.+ \tilde{m}_h^4 \ln (\tilde{m}_h^2/m_h^2)
  -12\tilde{m}_t^4 \ln (\tilde{m}_t^2/m_t^2)~\right],
  \label{VCW}
}
where
\al{
C_0 &\simeq -\frac{1}{64 \pi^2 v_h^4}\left(3m_W^4+(3/2)m_Z^4+(3/4) m_h^4-6 m_t^4\right),\label{C0}\\
\tilde{m}_W^2 &=(m_W/v_h)^2h^2,~
\tilde{m}_Z^2=(m_Z/v_h)^2 h^2,~
\tilde{m}_t^2 =(m_t/v_h)^2 h^2,\nn
\tilde{m}_h^2 &=
3\lambda_H h^2
+\frac{\lambda_{HS}}{64\pi^2}\bigg\{7 N_c N_f \lambda_{HS}h^2
-4 f N_c N_f (N_f\lambda_{S}+\lambda'_{S})\nn
&\quad -2 N_c N_f \left[-3 \lambda_{HS} h^2
+4f (N_f \lambda_S+\lambda'_S)\right]
  \ln \frac{4 f (N_f\lambda_S+\lambda'_S)
-\lambda_{HS}h^2}{2 \Lambda_H^2}\bigg\},
\label{mh2}
}
and $m_I$ ( masses given at the vacuum $v_h=246$ GeV) are 
\al{
m_W&=80.4~\mbox{GeV},& m_Z&=91.2~\mbox{GeV}, &m_t&=174~\mbox{GeV},& m_h&=125~\mbox{GeV}.&
\label{mass at vacuum v}
}
We find that the Coleman--Weinberg potential  \eqref{VCW} yields a one-loop correction 
to the Higgs mass  squared \eqref{vev2} 
\al{
\delta m_h^2 = \frac{d^2V_\text{CW}}{dh^2}\bigg|_{h=v_h}
 \simeq -16 C_0 v_h^2,
\label{correction to higgs}
}
which gives about $-6$\% correction to  \eqref{vev2}.

\section{Benchmark points and scale phase transition}\label{explicit point}
\begin{figure}
 \includegraphics[width=6cm]{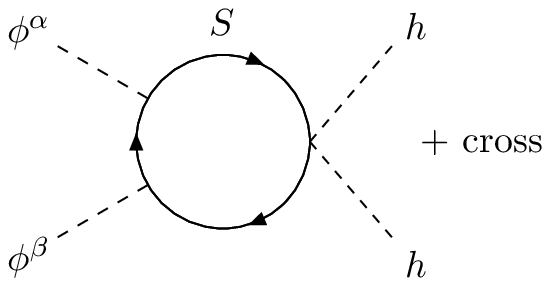}
  \includegraphics[width=12cm]{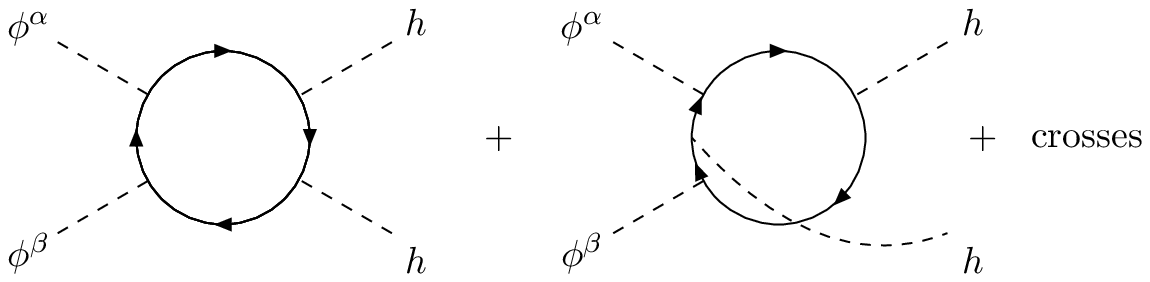}
\caption{
The effective interaction $\phi^2h^2$ generated by the $S$ loop effect.
}
\label{vertex}
\end{figure}

Thanks to the unbroken $U(N_f)$ flavor symmetry
the excitation field $\phi^a$ can be a DM candidate.
How  to evaluate  its relic abundance $\Omega_\text{DM} \hat{h}^2$ and its spin-independent elastic cross section off the nucleon $\sigma_{SI}$ is explained in \cite{Kubo:2015cna}, where $\hat{h}$ is the dimensionless Hubble constant today.
Since we would like to perform a benchmark point analysis,
let us explain the parameter space which we are interested in.
To obtain the DM relic abundance and its spin-independent elastic cross section off the nucleon, the annihilation process, in which
a pair of $\phi^a$ annihilates into the SM particles
through  the effective interaction $(\phi^a)^2 h^2$, has to be considered.
 The effective interaction 
  is generated by the loop effects of $S$ as shown in Fig.~\ref{vertex},
  where the vertices in the diagrams are 
   $S^\dag_i t^a_{ij} \phi^a S_j$ and $(S^\dagger_i S_i)h^2$ in the mean-field Lagrangian \eqref{MFA3}.
 The s-channel  describes the DM annihilation process, while the t-channel 
 is used for
 the DM interaction with the nucleon.
 
\begin{figure}
\includegraphics[width=10cm]{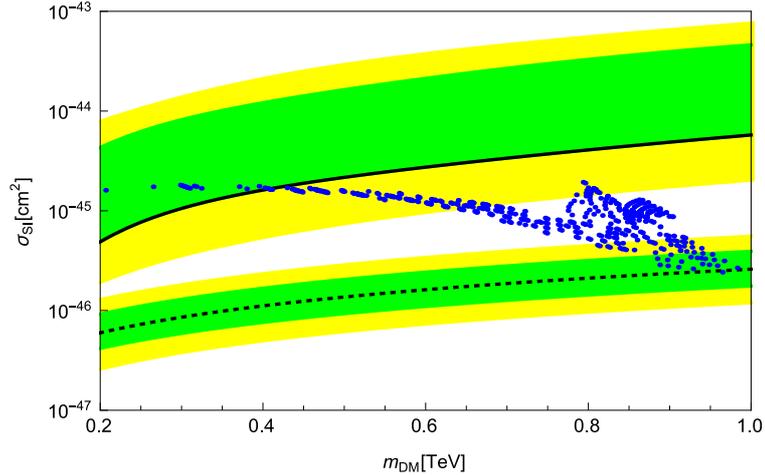} 
\caption{
The spin-independent elastic cross section $\sigma_{SI}$ 
of DM off   the nucleon as a function of $m_{\rm DM}$ for $N_f=2,N_c=6$.
The black solid line stands for the central value of the LUX upper bound
\cite{Akerib:2016vxi} with  one (green) and  two (yellow) $\sigma$ bands,
and the black dotted line indicates the sensitivity of
 XENON1T~\cite{Aprile:2012zx,Aprile:2015uzo}. }
\label{mDMsigma}
\end{figure}

 The Higgs portal coupling $\lambda_{HS}$ plays a triple role.
First, as we see from \eqref{vev1}, the smaller $\lambda_{HS}$ is, 
the larger $\Lambda_H$ has to become,
because $|\langle H\rangle|$ is fixed at $v_h/\sqrt{2}=246/\sqrt{2}$ GeV.
Secondly, since the coupling constant 
for the effective interaction  $(\phi^a)^2 h^2$ is
proportional to $\lambda_{HS}$, the DM relic abundance decreases
as $\lambda_{HS}$ increases, while $\sigma_{SI}$ increases.
That is,
to satisfy the experimental constraints on 
$\Omega_\text{DM}  \hat{h}^2$ \cite{Planck:2015xua} and 
$\sigma_{SI}$     \cite{Akerib:2013tjd} at the same time,
 $\lambda_{HS}$ has to lie in an interval.
 Equivalently, too small or too large $\Lambda_H$ is inconsistent
 with the DM constraints. 
 The interval depends strongly on $N_f$, because 
$\Omega_\text{DM}  \hat{h}^2$ is proportional to $N_f^2-1$. This implies that the weakest constraint on $\lambda_{HS}$
is given for $N_f=2$. Furthermore, the color degrees of freedom
in the hidden sector is not completely free within our
effective field theory approach. 
This is due to the inverse propagator $\Gamma(p^2)$
for the DM, which can have a zero for 
a positive $p^2$
only if  $\lambda'_S N_c$ is  large enough
(the zero of $\Gamma(p^2)$ defines the DM mass).
If we restrict ourselves to
$\lambda'_S \lsim 2$, we find that $N_c > 4$.
On the other hand, the results (at least for the DM phenomenology) 
for $N_c=5-8$ are very similar.
The predicted region for $N_f=2$ and $N_c=6$ is shown in Fig.~\ref{mDMsigma}. 
As we can see from 
Fig.~\ref{mDMsigma} this result could be  tested by XENON1T~\cite{Aprile:2012zx,Aprile:2015uzo}, whose sensitivity is indicated by the dotted line.

When discussing the scale phase transition at finite temperature
and the GW background,
we will consider a benchmark point in our parameter space:
\al{
N_f&=2,&
N_c&=6,&
\lambda_S&=0.145,&
\lambda'_S&=2.045,&
\lambda_H&=0.15,&
\lambda_{HS}&=0.032,&
\label{bench}
}
which yields 
\al{
\Lambda_H&=  0.0621~\text{TeV},&
m_\text{DM}&= 0.856~\text{TeV},&
\Omega_\text{DM}  \hat{h}^2&=0.122,&\nn
m_h&=0.126~\text{TeV},&
\sigma_{SI}&= 5.12\times10^{-46}~\text{cm}^2.&
}
As we can see from Fig.~\ref{mDMsigma},
most of the  predicted points in the 
$m_{\rm DM}-\sigma_{SI}$ plane,
except for those
with smaller $m_{\rm DM}$,  are close to the benchmark point.
The third role of $\lambda_{HS}$ will be discussed
when considering phase transition at finite temperature below.

It is expected that at high temperature the thermal effects restore the electroweak symmetry and scale symmetry. 
We  assume that  even at finite temperature 
the mean-field approximation is still a good
approximation to the original strongly interacting gauge theory
\eqref{LH}.
The effective potential consists of four components,
\al{
V_\text{eff}\fn{f,h,T} = V_\text{MFA}\fn{f,h} + V_\text{CW}\fn{h} + V_\text{FT}\fn{f,h,T} + V_\text{ring}\fn{h,T},
\label{thermal effective potentials}
}
where $ V_\text{MFA}\fn{f,h}$ and $V_\text{CW}\fn{h}$ are 
given in \eqref{effective potential MFAa} and \eqref{VCW}, respectively.
Since  the absolute minimum of $V_\text{MFA}$ is located at $\bra \bar S \ket =0$, we suppress the $\bar S$ dependence of $V_\text{MFA}$.
The main thermal effects are included in $V_\text{FT}\fn{f,h,T}$, which is 
\al{
V_\text{FT}\fn{f,h,T} &=  \sum_{I=S,W,Z,h}n_I T \int \frac{d^3k}{(2\pi)^3} \ln\fn{1+e^{\omega_I/T}} - n_tT\int \frac{d^3k}{(2\pi)^3} \ln\fn{1-e^{\omega_
t/T}}\nn
&=\frac{T^4}{2\pi^2}\bigg[
2N_cN_f  J_B\fn{\tilde M^2\fn{T}/T^2} 
+J_B\fn{\tilde m_h^2\fn{T}/T^2} 
+6J_B\fn{\tilde m_W^2/T^2} \nn
&\quad + 3J_B\fn{\tilde m_Z^2/T^2} 
-12 J_F\fn{\tilde m_t^2/T^2} 
\bigg],
\label{finite temperature potential}
}
where $\omega_I=\sqrt{\vec k^2 + m_I^2\fn{h}}$, and the thermal masses are 
given by
\al{
\tilde M^2\fn{T} &= M^2 + \frac{T^2}{6}\bigg( (N_cN_f +1)\lambda_S + (N_f+N_c)\lambda'_S -\lambda_{HS} \bigg),\\
\tilde m_h^2\fn{T}&= \tilde m_h^2 +\frac{T^4}{12} \bigg( \frac{9}{4}g_2^2+\frac{3}{4}g_1^2 +3y_t^2 +6\lambda_H -N_cN_f \lambda_{HS} \bigg).
}
Here $g_2=0.65$ and $g_1=0.36$ are the $SU(2)_L$ and $U(1)_Y$ gauge coupling constants,
respectively, while $y_t=1.0$ stands for
the top-Yukawa coupling constant.
Further, $M^2$ and $\tilde m_I (I=W,Z,t,h)$ are given in \eqref{cons scala mass}
 (with $H^\dagger H=h^2/2$) and \eqref{mh2}, respectively. The thermal functions and their high temperature expansions are 
\al{
J_B\fn{r^2} &=\int_0^\infty dx\, x^2 \ln
\left(1-e^{-\sqrt{x^2+r^2} } \right)\nn
&\simeq 
-\frac{\pi^4}{45}+\frac{\pi^2}{12}r^2
-\frac{\pi}{6}r^{3}-\frac{r^4}{32}
\left[\ln (r^2 /16\pi^2)+2\gamma_E-\frac{3}{2}   \right]&
&\mbox{for}~r^2\lsim 2,&
\label{JB}\\
J_F\fn{r^2} &=\int_0^\infty dx\, x^2 
\ln\left(1+e^{-\sqrt{x^2+r^2} } \right)\nn
&\simeq
\frac{7\pi^4}{360}-\frac{\pi^2}{24}r^2
-\frac{r^4}{32}\left[\ln (r^2 /\pi^2)+2\gamma_E-\frac{3}{2}
   \right] 
   &\mbox{for}~r^2\lsim 2.&
   \label{JF}
}
Although the high temperature expansions are useful, they are not suitable for large $r>2$.
Therefore, the following fitting functions~\cite{Funakubo:2009eg} are used\footnote{The validness of the approximations for the thermal functions is discussed in \cite{Hamada:2016gux}.}:
\al{
J_{B(F)}(r^2)
&\simeq 
e^{-r}\sum_{n=0}^{40} c_n^{B(F)} r^n.
}
The contribution from the ring (daisy) diagrams of gauge boson is
\cite{Carrington:1991hz}
\al{
V_{\rm ring}\fn{h,T}&=
-\frac{T}{12\pi}\bigg(
2 a_g^{3/2}+\frac{1}{2\sqrt{2}}\left(a_g+c_g-[(a_g-c_g)^2+4 b_g^2]^{1/2}\right)^{3/2}\nn
&\quad+\frac{1}{2\sqrt{2}}\left(a_g+c_g+[(a_g-c_g)^2+4 b_g^2]^{1/2}\right)^{3/2}-\frac{1}{4}[g_2^2 h^2]^{3/2}
-\frac{1}{8}[(g_2^2+g_1^2) h^2]^{3/2}\bigg),
}
where
\al{
a_g &=\frac{1}{4}g_2^2 h^2+\frac{11}{6}g_2^2 T^2,&
b_g &= -\frac{1}{4}g_2 g_1 h^2,&
c_g &= \frac{1}{4}g_1^2 h^2+\frac{11}{6}g_1^2 T^2.&
}
Note that the ring contributions from the scalar $S$ and the Higgs 
field are included in \eqref{finite temperature potential}.

Using the thermal effective potential \eqref{thermal effective potentials}, 
it is possible to consider phase transitions at finite temperature:
There exist two phase transitions in this model,
 the scale and EW phase transitions (whose critical temperatures are denoted by $T_\text{S}$ and $T_\text{EW}$, respectively).
The scale phase transition in hidden sector can be strongly
first-order  for a wide range in
the parameter space \cite{Kubo:2015joa}. 
In the case with $N_f=1$ (no  DM) we see that both phase transitions can be strongly
first-order and occur at the same temperature, $T_\text{S}=T_\text{EW}$.
On the other hand, 
if DM is consistently included ($N_f > 1$),
the EW phase transition becomes weak, although the scale phase transition  is still strongly first-order, and  $T_\text{EW} < T_\text{S}$.
The crucial difference between the two cases comes from the value of the Higgs portal coupling constant $\lambda_{HS}$, which controls the strength
of the connection  between the SM and hidden sector. 
As discussed in section~\ref{explicit point},  smaller $\lambda_{HS}$ means
 larger $\Lambda_H$, which in turn implies  larger $T_S$,
while $T_\text{EW}$ stays around $O(100)$ GeV.
This is the third role of $\lambda_{HS}$.
Therefore, within the minimal model with a DM candidate 
we are considering here,
the scale phase transition occurs at a higher temperature 
than the EW phase transition and is strongly first-order.
Since we are especially interested in the possibility to explain the origin
of the EW scale and DM at the same time, we focus in the following
discussions on the model with $N_f=2$,
in which, as explained 
in section~\ref{explicit point},
 the weakest constraint on $\lambda_{HS}$ has to be satisfied. 

\begin{figure}
\includegraphics[width=10cm]{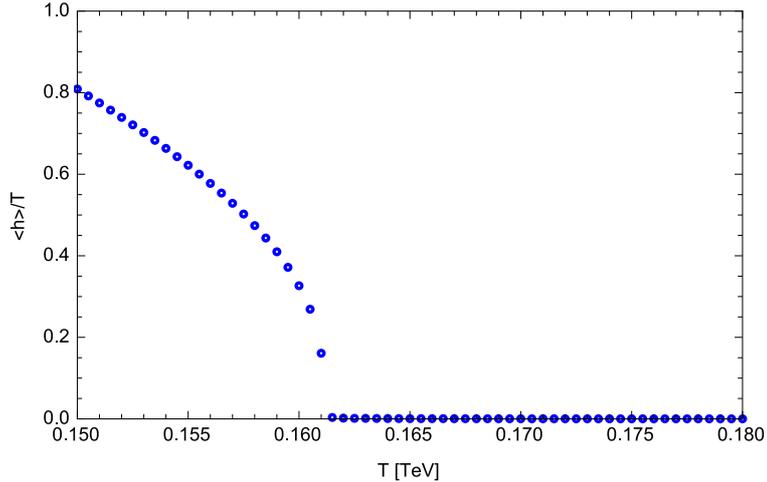} 
\caption{\label{EWPT}
The EW  phase transition.
The transition occurs around $T=T_\text{EW}\simeq 0.161$ TeV.
}
\end{figure}

\begin{figure}
\includegraphics[width=8cm]{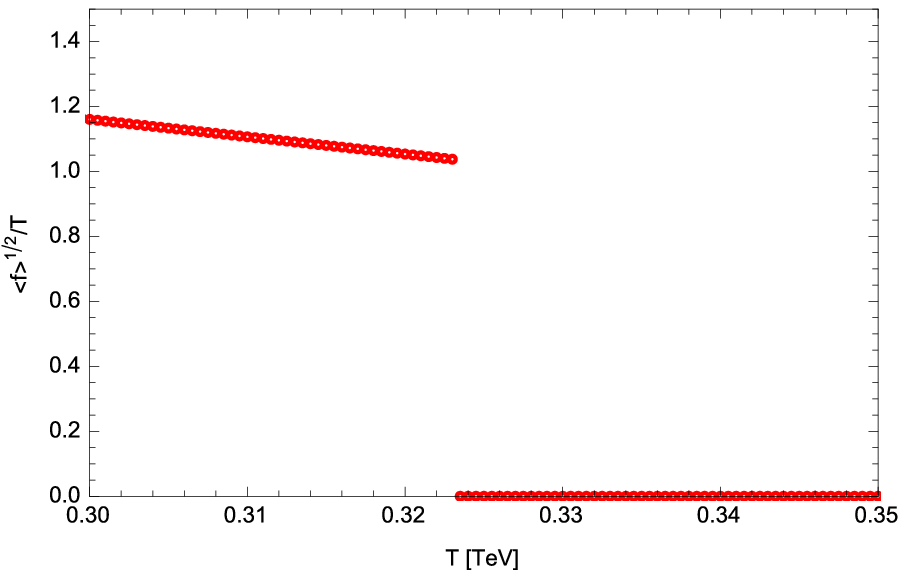} 
\includegraphics[width=8cm]{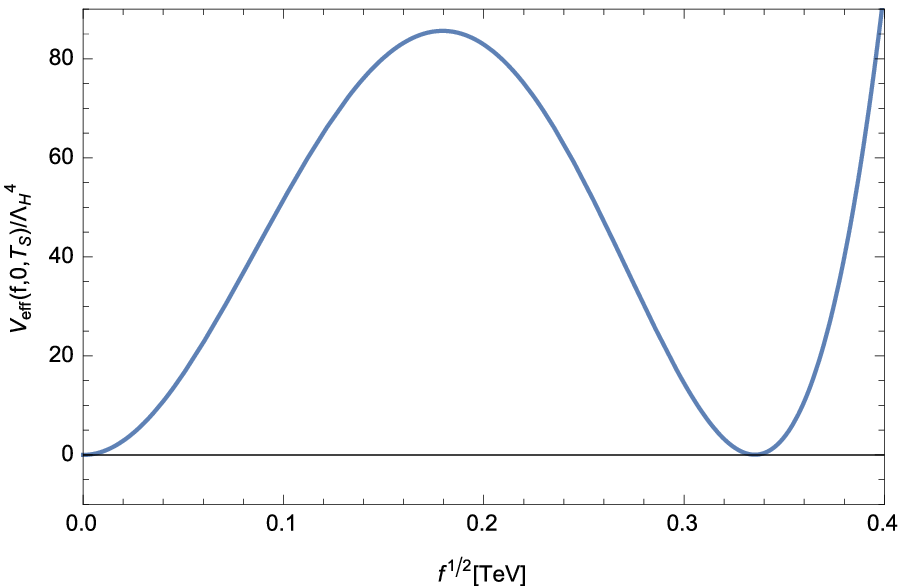}
\caption{\label{SPT}
 Left: The scale  phase transition.
The strong first-order  transition occurs at $T=T_{S}=0.323$ TeV.
Right: The effective  potential 
$V_{\rm eff}(f,0,T)/\Lambda_H^4$ against 
$f^{1/2}$ at the critical temperature
$T=T_{S}=0.323$ TeV. The potential energy density at the origin is subtracted 
from $V_{\rm eff}$.
}
\end{figure}

The EW phase transition is shown in Fig.~\ref{EWPT} for the set of the benchmark 
parameters \eqref{bench}. We see from Fig.~\ref{EWPT} 
that a weak  transition appears around $T_\text{EW}\simeq  0.161$ TeV.
Fig.~\ref{SPT} (left) presents $\langle f \rangle^{1/2}/T$  as a function of $T$,
while
Fig.~\ref{SPT} (right)
 shows the effective potential $V_\text{eff}$ given in \eqref{thermal effective potentials}
 with $h=0$ at $T=T_S=0.323$ TeV as a function of $f^{1/2}$, 
    showing that the scale phase transition is strongly first-order.
    
 
\section{Gravitational wave from scale phase transition}\label{GW analysis}

As we have seen in the previous section, the scale phase transition
occurs at $T\lsim$ few hundred GeV. This is  larger than
$T_\text{EW}$, and moreover the phase transition is strongly first-order\footnote{The GW background produced by a strong EW phase transition has been studied in 
\cite{Grojean:2004xa,Kehayias:2009tn,Leitao:2012tx,Kikuta:2014eja,Jinno:2015doa,Leitao:2015fmj,Kakizaki:2015wua,Jinno:2016knw}.
Similar studies have been made in \cite{Jaeckel:2016jlh,Dev:2016feu}.}.
This strong first-order phase transition produces the EW scale and at the same time   dark matter  in the early Universe.
Fortunately, this phase transition could be observed indirectly,
because it can generates stochastic
GW background, which can be detected
in GW experiments \cite{Witten:1984rs,Hogan:1986qda,Turner:1990rc}.
The most important quantity in studying the GW relics
is the GW background spectrum (see for instance \cite{Maggiore:1999vm,Binetruy:2012ze})
 $\Omega_\text{GW}\fn{\nu}\sim \frac{d\rho_\text{GW}}{d\log \nu}$, 
 where $\nu$ is the frequency of the GW, and $\rho_\text{GW}$ is   
 the energy density of the GW. The energy density
 $\rho_\text{GW}$ is proportional to $\bra \dot h_{ij} \dot h_{ij}\ket$,
 where $h_{ij}$ is the transverse, traceless, spatial
 tensor perturbation of  the Robertson--Walker metric, and 
the dot on $h_{ij}$ denotes the derivative with respect to time.
The evolution of  $h_{ij}$ is described by the Einstein equation with the energy-momentum tensor which contains the information about the phase transitions.

The characteristics of a first-oder phase transition
in discussing the GW  background spectrum are the duration of the phase transition
and the latent heat released.
The duration time has to be sufficiently short compared with the expansion of the 
Universe, and it is clear that the more latent heat is released, the larger
$\Omega_\text{GW}$ can become. In the following two subsections
we will discuss these issues.

\subsection{Latent heat}
Given  the effective potential $V_\text{eff}$ in 
\eqref{thermal effective potentials}, it is straightforward
to compute the latent heat $\epsilon(T)$ at $T < T_S$:
\al{
\epsilon\fn{T} = -V_\text{eff}\fn{f_B\fn{T},T} + T \frac{\p V_\text{eff}\fn{f_B\fn{T},T}}{\p T},
}
where $f_B\fn{T}$ is $\bra f \ket$ at $T$, and we have set $h$ equal to
zero, because $\bra h \ket=0$ for $ T > T_{EW}$.
 The ratio of the released vacuum density $\epsilon\fn{T_t}$ 
  to the radiation energy density is 
  \al{
\alpha =\frac{\epsilon\fn{T_t}}{\rho_\text{rad}^t},
\label{alpha}
}
which is one of the basic parameters
  entering into $\Omega_\text{GW}$,
where $T_t$   (defined in \eqref{transition}) is the 
temperature just below $T_S$,
and $\rho_\text{rad}^t= g_*\fn{T_t}\pi^2T^4_t/30$  with $g_*=106.75$.

Since the latent heat in the lattice  $SU(3)_c$ gauge theory 
has been calculated \cite{Shirogane:2016zbf},
it may be worthwhile to compare our result with that of 
\cite{Shirogane:2016zbf}.
To this end, we calculate the latent heat in the mean-field approximation
with $N_c=3, N_f=1$ and $ \lambda_{HS}=\lambda_{H}=0$.
In this case the dynamical  scale symmetry breaking  (at $T=0$) occurs for 
$\lambda_S+\lambda'_S \geq 2.8$, and therefore, we calculate
the latent heat $ \epsilon\fn{T}$ just below the critical temperature for 
$\lambda_S+\lambda'_S =3,4$ and $5$, respectively.
We find:
\al{
\frac{\epsilon\fn{T}}{T^4} &=
\begin{cases}
0.70\\
0.55 \\
0.43
\end{cases}&
& \mbox{for}&
\lambda_S+\lambda'_S&=
\begin{cases} 
3\\
4\\
5
\end{cases}.&
\label{LHQCD}
}
These results should be compared with $\epsilon\fn{T}/T^4=
0.75\pm 0.17$ of \cite{Shirogane:2016zbf}. 
Though this lattice result is
obtained in the theory  without matter field, we see that the values
\eqref{LHQCD} are comparable in size to the lattice result.
This is a good news, because 
the scale phase transition and the deconfinement phase transition
appear at the same time, and the latent heat is proportional to the
change of entropy which we do not expect to change a lot if
one scalar field is  included or not.

\subsection{Duration time}
To estimate the duration time of a first-order phase transition,
we have to consider the underlying physical process, the tunneling process
from the false vacuum to the true vacuum,
because the duration time is the inverse of the decay rate
of the false vacuum.  
The  decay (the tunneling probability) per unit time per unit volume 
$\Gamma(t)$
 can be written as
\al{
\Gamma\fn{t} \sim  e^{-S_E\fn{t}},
}
where $S_E $ is the  Euclidean action
 in the full theory \eqref{LH}. At finite temperature $T$ the theory is
 equivalent to  the   Euclidean theory, which is periodic
in the  Euclidean time with the period of $T^{-1}$.
Above a certain high temperature
the typical size of the bubbles generated by the phase transition
may become much larger than the period $T^{-1}$ \cite{Linde:1981zj}.
Then we may assume \cite{Linde:1981zj} that 
 $S_E =S_3/T$, where $S_3$ is 
the corresponding three-dimensional action.

There are various complications when computing $S_3$ in the effective theory
in the mean-field approximation, where they are related with each other.
First,  the canonical dimension of the mean field $f$ is two.
This itself  is not a big problem, and we redefine $f$ as
\al{
f = \gamma ~\chi^2,
\label{chi}
}
where $\gamma$ is dimensionless so that the canonical dimension of 
$\chi$ is one.
Second, the kinetic term for $f$ is absent at the tree-level and 
is generated in  the one-loop order. Consequently, the kinetic term
of $f$ in the action involves the field renormalization factor
$Z$. Since the effective potential is computed at zero external momenta,
$Z$ may be computed also at zero external 
momenta, which is done  in the appendix.
As we can see from \eqref{WFRZ}, $Z$ depends on $f$ as well as on $T$.
However, we have a problem here, because how to include $Z$
in the kinetic term  is not unique:
$Z^{-1}(\partial_i f)(\partial_i f)$ and 
$(\partial_i Z^{-1/2}f)(\partial_i Z^{-1/2}f)$, for instance,
give different equations of motion for $f$.
The third complication is most serious:
In principle we have to compute
$S_E$ in the full theory \eqref{LH}. That is, 
instantons have to be taken into account. Therefore, we expect that the
action  $S_3/T$ computed in the effective theory in
the mean-field approximation can  not be a good approximation to 
the full quantity, because the effective theory does not know 
about instantons and confinement.
So, we do not trust $S_3/T$ obtained in the effective theory,
although we believe that the effective potential $V_\text{eff}$ 
is a good approximation (in fact the latent heat computed 
from $V_\text{eff}$ gives reasonable values compared with the lattice 
result \cite{Shirogane:2016zbf}). 
That is, we do not trust the kinetic term obtained
in the effective theory
(which is anyhow ambiguous), and instead we make an Ansatz
for the kinetic term. The simplest assumption is that the redefinition
\eqref{chi} gives a correct, canonically normalized kinetic term for $\chi$,
where $\gamma$ should be regarded as a free parameter constant independent
of $\chi$ and $T$.\footnote{As we see from \eqref{WFRZero}, 
$Z$ at $T=0$, obtained in the one-loop order in the effective theory,
is $16 \pi^2 12 N_f (N_f\lambda_S+\lambda'_S) f
\sim 8\times 10^2 f$ for the benchmark parameters \eqref{bench}.
Then $Z^{-1}\partial_i f\partial_i f=\partial_i \chi\partial_i \chi$
if $\gamma=4 \pi^2 12 N_f (N_f\lambda_S+\lambda'_S)\sim 2\times 10^2$.
This implies that, although $\bra f^{1/2} \ket /T_S\sim 1$ 
for the benchmark point  \eqref{bench},
$\bra \chi \ket /T_S\sim 0.02$ for  $\chi$. 
So, in terms of $\chi$ the phase transition
would no longer be a strong phase transition. This is also a reason why we do not
trust the kinetic term for $f$ obtained in the effective theory.}
That is, we make an Ansatz for $S_3$:
\al{
S_3\fn{T}=\int d^3x \left[ \frac{1}{2}(\nabla_i \chi)^2 +
V_\text{eff}\fn{\gamma \chi^2,T}\right],
\label{S3}
}
where the $h$ dependence of the effective potential is suppressed
(because $\bra h \ket =0$ for $ 
T > T_\text{EW}$), and  it is normalized as 
$V_\text{eff}\fn{0,T}=0$. 
At high enough temperatures we may assume \cite{Linde:1981zj} 
 that, not only
$S_E=S_3/T$ is satisfied, but also the classical solution to the field equation
is $O(3)$ symmetric: $\chi$ depends only on $r=(x_1^2+x_2^2+x_3^2)^{1/2}$
and satisfies
\al{
\frac{d^2\chi}{dr^2} + \frac{2}{r}\frac{d\chi}{dr} = 
\frac{dV_\text{eff}\fn{\gamma\chi^2,T}}{d\chi}.
\label{EoMbounce}
}
To compute the tunnel provability at $T < T_S$ 
from the false vacuum with $\chi=0$ to the true vacuum
with $\chi=\chi_B\neq 0$, we look for the classical solution,
the so-called bounce solution, that satisfies 
the boundary conditions, $d\chi/dr|_{r=0}=0$ and $\chi\fn{r=\infty}=0$.
The initial value  $\chi(0)$ should be 
 chosen slightly smaller than $\chi_B$, such that 
$\chi\fn{r=\infty}=0$ is satisfied. Then we insert the solution into 
\eqref{S3} to compute $S_3(T)/T$.

At the temperature $T_t$ 
 (or at the time $t_t$) the vacuum is overwhelmed by the bubble of the broken phase.
 Since the Universe is expanding, it is the time at which
 the tunneling provability per Hubble time per Hubble volume
 becomes nearly one,
 because after each  tunneling process we have one bubble nucleation.
This defines the transition time $t_t$ and also the transition temperature $T_t$:
\al{
\frac{\Gamma\fn{t}}{H(t)^4}\bigg|_{t=t_t}\simeq 1,
\label{transition}
}
where $H(t)$ is the Hubble parameter at $t$.
This condition is rewritten as \cite{Kehayias:2009tn}
\al{
S_E\fn{t_t}=\frac{S_3\fn{T_t}}{T_t}\simeq 140\text{--}150.
\label{transition1}
}
Since the transition time $t_t$ (or the transition temperature
$T_t$) is now defined, we can compute the duration of the phase
transition.  To this end, we expand the action around $t_t$:
\al{
S_E\fn{t}=S_E\fn{t_t}-\beta \Delta t+O((\Delta t)^2),
}
where $\Delta t=(t-t_t)>0$.
Then the tunneling per unit time per volume
can be approximated as 
\al{
\Gamma(t)\sim e^{\beta \Delta t}.
}
Therefore,  $\beta^{-1}$  is  the duration time and
can be computed from
\al{
\beta &= -\frac{dS_E}{dt}\bigg|_{t=t_t} =\frac{1}{\Gamma}\frac{d \Gamma}{dt}\bigg|_{t=t_t}
= H_t T_t\frac{d}{dT} \left( \frac{S_3\fn{T}}{T} \right)\bigg|_{T=T_t}\nn
& =H_t \tilde{\beta},
\label{beta}
}
where $d T/d t=-H T$ is used and $H_t$ is $H(t_t)$.

This $\tilde{\beta}$ and $\alpha$ in \eqref{alpha}
are the parameters which determine the characteristics
of the first-order phase transition and enter into the GW background spectrum.
Their values for  the set of the benchmark parameters \eqref{bench} 
with $\gamma=0.5, 1$ and $2$ are shown in Table \ref{list}.

\subsection{GW background spectrum}
It is known at present that there are three production mechanisms of  GWs at 
a strong first-oder phase transition:
The bubble nucleation in a strong first-order phase  transition
 grows, leading to the contributions to $\Omega_\text{GW}$
from collisions of bubble walls  $\Omega_\text{coll} $  
\cite{Kosowsky:1991ua,Kosowsky:1992rz,Kosowsky:1992vn,Kamionkowski:1993fg,Caprini:2007xq,Huber:2008hg,Jinno:2016vai},
magnetohydrodynamic  turbulence $\Omega_\text{MHD} $
\cite{Kosowsky:2001xp,Dolgov:2002ra,Caprini:2006jb,Gogoberidze:2007an,Kahniashvili:2008pe,Caprini:2009yp,Kisslinger:2015hua}
and also
sound waves $\Omega_\text{sw}$
after the bubble wall collisions \cite{Hindmarsh:2013xza,Hindmarsh:2015qta,Giblin:2013kea,Giblin:2014qia}.
Then the total GW background  spectrum is given by
\al{
\Omega_\text{GW} (\nu)\hat{h}^2 = 
\left[\Omega_\text{coll}(\nu) +
\Omega_\text{MHD} (\nu) +\Omega_\text{sw}(\nu)\right] \hat{h}^2,
}
where $\hat{h}^2$ is the dimensionless Hubble parameter 
and $\nu$ is the frequency of the GW at present.
The individual contributions to the GW background  spectrum
 can be estimated for given $\alpha$
and $\tilde{\beta}$ along with the bubble wall velocity $v_b$ and
the efficiency factor $\kappa$.  It is the fraction of 
the liberated vacuum energy
 to the bulk kinetic energy of the fluid, 
 which can become the source of the GW.
 \begin{table}
\begin{center}
\begin{tabular}{|c|c|c|c|c|c|c|} 
\hline
$\gamma$ & $T_t$ [TeV] & $S_3(T_t)/T_t $ & $\alpha$ & $\tilde{\beta}$ & 
$\tilde{\Omega}_\text{sw}\hat{h}^2 $ & $\tilde \nu_\text{sw} $ [Hz] \\ \hline
$0.5$ & $0.300$ & $149$ & $0.070$ & $3.7\times10^3$ & $1.9\times 10^{-13}$ &
$0.37$ \\ \hline
$1.0$ & $0.311$ & $145$ & $0.062$ & $7.0\times 10^3$ &
 $7.4\times 10^{-14}$ &
$0.73$ \\ \hline
$2.0$ & $0.316$ & $146$ & $0.059$ & $13\times10^3$ & $3.4\times 10^{-14}$ &
$1.4$ \\ \hline
\end{tabular}
\caption{Relevant quantities for the GW background  spectrum 
for the set of the benchmark parameters \eqref{bench}.
The quantities $\alpha,\tilde{\beta}$ and $\gamma$ 
are defined in \eqref{alpha}, \eqref{beta} and \eqref{chi}, respectively.}
\label{list}
\end{center}
\end{table}

\begin{figure}
\includegraphics[width=11cm]{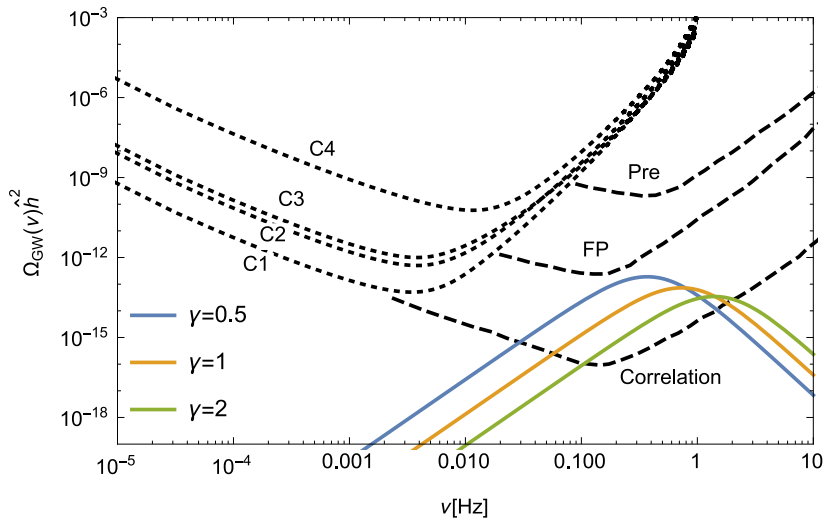} 
\caption{
The GW background spectrum.
The doted lines are the four different sensitivities of eLISA, where the labels (``C1", $\cdots$,``C4") correspond to the configurations listed in Table~1 in \cite{Caprini:2015zlo}.
The data sets of their configurations are taken from \cite{eLISAdata}.
The dashed lines are sensitivities of three different designs (``Pre-DECIGO", ``FP-DECIGO" and ``Correlation") of DECIGO~\cite{Seto:2001qf}.
The parameter $\gamma$ is defined in \eqref{chi}.
}
\label{spectrum}
\end{figure}
Using the formulas given in \cite{Caprini:2015zlo}, 
we have computed these individual 
contributions to the GW background 
spectrum 
for the set of the benchmark parameters \eqref{bench}
and found that
$\Omega_\text{coll}$ and
$\Omega_\text{MHD}$ are several orders of magnitude
smaller than
$\Omega_\text{sw}$.  Therefore, we consider here only
the contribution to $\Omega_\text{GW}$ from the sound wave
\cite{Hindmarsh:2013xza,Hindmarsh:2015qta,Giblin:2013kea,Giblin:2014qia}:
\al{
\Omega_\text{sw}\fn{\nu}\hat{h}^2=\tilde \Omega_\text{sw}
\hat{h}^2\left(\frac{\nu}{\tilde \nu_\text{sw}}\right)^3\left( \frac{7}{4+3\left(\nu/\tilde \nu_\text{sw}\right)^2} \right)^{7/2},
}
where
\al{
\tilde \Omega_\text{sw}\hat{h}^2=
2.65\times 10^{-6}v_b\tilde \beta^{-1} \left( \frac{\kappa \alpha}{1+\alpha} \right)^2
\left( \frac{100}{g_*}\right)^{1/3}
}
with the peak frequency $\tilde{\nu}_\text{sw}$ given by
\al{
\tilde \nu_\text{sw}=
1.9\times 10^{-5}\, \text{Hz}\times \frac{\tilde \beta}{v_b}
 \left( \frac{T_*}{100\,\text{GeV}} \right) \left(\frac{g_*}{100} \right)^{1/6}.
 }
Since we are interested in order-of-magnitude estimates
 of the GW background spectrum, we consider the case that the wall velocity
 is the sound velocity $c_s=0.577$. In this case the efficiency factor
 $\kappa$  becomes \cite{Hindmarsh:2013xza,Hindmarsh:2015qta,Giblin:2013kea,Giblin:2014qia}:
\al{
\kappa=\frac{\alpha^{2/5}}{0.017+(0.997+\alpha)^{2/5}}.
}
The result for the set of the benchmark parameters \eqref{bench}
with $\gamma=0.5, 1$ and $2$
is shown in Table I, and the model prediction of the GW background spectrum
is compared with the experimental sensitivity of present and future
experiments in Fig.~\ref{spectrum}.
As we can see from Fig.~\ref{spectrum},
the GW background produced by the scale phase transition
in the early Universe can be observed by DECIGO,
where $\gamma$ is introduced in \eqref{chi}
and its double meaning is explained there.

\section{Summary}\label{summary}
In this paper we have considered   a non-perturbative effect,
the gauge-invariant scalar-bilinear condensation, that 
generates a scale in a strongly interacting
 gauge sector, where the scale  is transmitted to the SM sector
via a Higgs portal coupling $\lambda_{HS}$.
The dynamical scale genesis appears as a phase transition
at finite temperature, and it can  produce
the GW background in an early Universe.
Since  $\lambda_{HS}$ controls the strength
of the connection  between the SM and hidden sector,
 smaller $\lambda_{HS}$ means
 larger $\Lambda_H$, which in turn implies larger critical temperature
$T_S$ for the scale phase transition. Therefore, 
$T_S\gg T_\text{EW}\sim O(100)$ GeV for   smaller $\lambda_{HS}$.
The coupling $\lambda_{HS}$, on the other hand, cannot be too small, because it is constrained from below
by the DM relic abundance, implying that $T_S < $ few hundreds GeV.
Interestingly, in this interval of $\lambda_{HS}$, where we obtain
consistent values of the DM relic abundance, the scale phase transition
is strongly first-order.

We have calculated the spectrum of  the  GW background,
using the effective field theory, which has been developed in 
\cite{Kubo:2015cna}.
Our intension in the present paper  has not been to describe in detail the 
process of the  generation of the GW background.
We instead have applied the formulas given in \cite{Caprini:2015zlo}
to compute the GW background spectrum.
We have found that the contributions to the GW spectrum
from the collisions of bubble walls and also
from the magnetohydrodynamic  turbulence 
are negligibly small compared with  the sound wave contribution.
We have found that the peak frequency $\tilde{\nu}_\text{sw}$ of 
the GW background is $O(10^{-1})\sim O(1)$ Hz with
the peak relic energy density $\Omega_\text{sw} \hat{h}^2 =O(10^{-14})
\sim O(10^{-13})$.
Therefore,
the scale phase transition caused by the scalar-bilinear
condensation is strong enough that 
the  corresponding GW background   
can be observed by DECIGO \cite{Seto:2001qf} in future.

\subsection*{Acknowledgements}
We would like to thank 
K.~Hashino, M.~Kakizaki and S.~Kanemura  for useful discussions
and suggestions.
J.~K.~is partially supported by the Grant-in-Aid for Scientific Research (C) from the Japan Society for Promotion of Science (Grant No.16K05315).
M.~Y.~is partially supported by the Grant-in-Aid for Encouragement of Young Scientists (B) from the Japan Society for Promotion of Science (Principal Investigator: Koji Tsumura, Grant No.16K17697).

\begin{appendix}
\section{Field renormalization factor}\label{kinetic terms}
We derive the kinetic term of the composite field at the  one-loop level.
The two-point function is given by
\al{
\Pi\fn{\omega_n,{\vec p},T, M}=T\sum_l \int\frac{d ^3k}{(2\pi)^3}
\frac{1}{(\omega_n-\omega_l)^2+(\vec p-\vec k)^2+M^2}
\frac{1}{\omega_l^2+\vec k^2+M^2},
\label{two point function}
}
where $M^2$ is \eqref{cons scala mass} with $\lambda_{HS}=0$.
Although at finite temperature the kinetic term between the time-direction momentum mode and spacial-direction one is anisotropic, hence the field renormalization factors should be defined as\footnote{More precisely, the field renormalization factor should be defined at on-shell external momentum. However, we calculate it at $p=0$ for simplicity.} 
\al{
Z^{-1}_\perp\fn{f}&\equiv \frac{d \Gamma}{d \omega_n^2}\bigg|_{p=0},&
Z^{-1}_\parallel\fn{f} &\equiv \frac{d \Gamma}{d \vec p^2}\bigg|_{p=0},&
}
we hereafter assume that  $Z^{-1}_\perp\approx Z^{-1}_\parallel$ and write them as simply $Z^{-1}$.

Expanding the integrant of \eqref{two point function} into the polynomial of $\vec p$ around $\omega_n=\vec p=0$,
\al{
&\frac{1}{(\omega_n-\omega_l)^2+(\vec p-\vec k)^2+M^2}\frac{1}{\omega_l^2+\vec k^2+M^2}\nn
&\quad =
\frac{1}{(\omega_l^2+\vec k^2+M^2)^2}
-\frac{\vec p^2 -2\vec p\cdot \vec k}{(\omega_l^2+\vec k^2+M^2)^3}
+\frac{4(\vec k\cdot \vec p)^2}{(\omega_l^2+\vec k^2+M^2)^4}+\cdots,
}
and replacing  $\vec p\cdot \vec k$ and its squared 
by $0$ and $\vec k^2 \vec p^2/3$, respectively, we obtain
\al{
Z^{-1}\fn{f}=
T\sum_{l}\int \frac{d^3 k}{(2\pi)^3}
\left(
\frac{1}{(\omega_l^2+\vec k^2+M^2)^3}
-\frac{4\vec k^2}{3(\omega_l^2+\vec k^2+M^2)^4}
\right).
}
Using the standard calculation method at finite temperature, we can separate this into the zero temperature mode and the finite one:
\al{
Z^{-1}\fn{f}=Z_{T=0}^{-1}\fn{f} + Z_{T\neq 0}^{-1}\fn{f},
}
with
\al{\label{WFRZero}
Z_{T=0}^{-1}\fn{f}= \int \frac{d^4 k}{(2\pi)^3}
\left(
\frac{1}{(k^2+M^2)^3}
-\frac{k^2}{(k^2+M^2)^4}
\right)
=\frac{1}{16\pi^2}\frac{1}{6M^2},
}
and
\al{
Z_{T\neq 0}^{-1}\fn{f} &= \int \frac{d^3 k}{(2\pi)^3}
\left(
\frac{3}{8}\frac{1}{(k^2+M^2)^{5/2}} +\frac{5}{12} \frac{1}{(k^2+M^2)^{7/2}}
\right)\frac{1}{e^{\sqrt{k^2+M^2}/T}-1} \nn
&= \frac{1}{16\pi^2T^2} \left( 3g_{3/5}\fn{M/T} + \frac{10}{3}g_{5/7}\fn{M/T} \right).
}
The thermal function is defined as
\al{
g_{n/m}\fn{y}= \int^\infty_0 d x\,  \frac{x^{n-1}}{(x^2+y^2)^{m/2}}\left( e^{\sqrt{x^2+y^2}} -1 \right)^{-1}.
\label{WFRZ}
}
\end{appendix}

\end{document}